\documentclass[a4paper, 11pt]{article}
\usepackage[utf8]{inputenc}
\usepackage[english]{babel}
\usepackage{amsmath,  amssymb, amsfonts}
\usepackage{amsthm}
\usepackage{mathrsfs}
\usepackage{color}
\usepackage{graphicx}
\usepackage{enumerate}
\usepackage{bm}
\usepackage{cite}
\usepackage{url}
\usepackage{float}
\usepackage{algorithm} 
\usepackage{algorithmic} 

\newcommand*\mcapinn[2]{\vcenter{\hbox{$\mathsurround=0pt
  \ifx\displaystyle#1\textstyle\else#1\fi\bigcap$}}}

\newcommand*\mcupinn[2]{\vcenter{\hbox{$\mathsurround=0pt
  \ifx\displaystyle#1\textstyle\else#1\fi\bigcup$}}}

\DeclareFontFamily{OT1}{pzc}{}

\DeclareFontShape{OT1}{pzc}{m}{it}{<-> s * [1.200] pzcmi7t}{}

\DeclareMathAlphabet{\mathpzc}{OT1}{pzc}{m}{it}

\newtheorem{theorem}{Theorem}

\title{\LARGE \bf Quantum  Clique Gossiping}

\author{Bo Li\thanks{B. Li  is with the Key Laboratory of Mathematics Mechanization (KLMM), Academy of Mathematics and Systems Science, Chinese Academy of Sciences, Beijing 100190, China. Email:  libo@amss.ac.cn.}, Shuang Li\thanks{S. Li  is with the Key Laboratory of Mathematics Mechanization (KLMM), Academy of Mathematics and Systems Science, University of  Chinese Academy of Sciences, Beijing 100190, China. Email:  lishuang16@mails.ucas.ac.cn.}, Junfeng Wu\thanks{J. Wu is with the College of Control Science and Engineering, Zhejiang Universiy, Hangzhou 310027, China. Email: jfwu@zju.edu.cn.}, and Hongsheng Qi\thanks{H. Qi is with the Key Laboratory of Systems and Control, Academy of Mathematics and Systems Science, Chinese Academy of Sciences, Beijing 100190, and the University of Chinese Academy of Sciences, Beijing 100049, P. R. China. Email: qihongsh@amss.ac.cn.}
  }
\date{}

\begin{document}

\maketitle

\begin{abstract}
This paper establishes a framework for the acceleration of quantum gossip algorithms by introducing  local clique operations to networks of interconnected qubits. Cliques are local structures in complex networks being complete subgraphs. Based on cyclic permutations, clique gossiping leads to collective multi-party qubit interactions. This type of algorithm can be physically realized by a series of local environments using coherent methods. First of all, we show that at reduced states, these cliques have the same acceleration effects as their roles in accelerating classical gossip algorithms, which can even make possible  finite-time convergence for suitable network structures. Next, for randomized selection of cliques where  node updates enjoy a more self-organized and scalable sequencing, we show that the rate of convergence is precisely improved by $\mathcal{O}(k/n)$ at the reduced states, where $k$ is the size of the cliques and $n$ is the number of qubits in the network.  The rate of convergence at the coherent states of the overall quantum network is proven  to be decided by  the spectrum of a mean-square error evolution matrix. Explicit calculation of such matrix is rather challenging, nonetheless, the effect of cliques on the coherent states' dynamics is illustrated via numerical examples. Interestingly, the use of larger quantum cliques does not necessarily increase the speed of  the network density aggregation, suggesting quantum network dynamics is not entirely decided by its  classical topology.  \end{abstract}

\section{Introduction}

Recently the study of distributed quantum consensus algorithms drew attention in the research community \cite{Ticozzi,Shi-TAC,tac-2017,sr-2017}, where the goal is to build the analogue of classical consensus algorithms \cite{Pease:1980:RAP:322186.322188,tsitsiklis,Jadbabaie2003,murray} towards distributed and scalable control and computation means for quantum networks. In classical networks, nodes holding real values can achieve a common state by self-organized communication and local computations \cite{Jadbabaie2003}. In networks of qubits, consensus can be defined over a set of different notions\cite{Ticozzi}, but using the idea of classical gossip algorithms\cite{Boyd2006} quantum consensus algorithms  can indeed be developed with conceptual consistency. For both open quantum networks\cite{Shi-TAC,tac-2017} and hybrid quantum networks\cite{sr-2017}, quantum consensus can in fact be conveniently studied by building the bridge to its classical counterpart.

The prospects of carrying out quantum computation and quantum communication via networks of quantum subsystems have already been noted in the past few years \cite{duan,cirac2007,PhysRevLett.103.240503,CiracRandomNetworks2011,dissipation}. The development of quantum consensus and synchronization algorithms\cite{Ticozzi,Shi-TAC,tac-2017,sr-2017} marks a continuing flow of this line of research as simple but fundamental blocks of network computation and information dissemination. The expectation is that  more advanced algorithms may be   developed for a variety of control, computation,  and estimation tasks over complex quantum  networks on top of such foundations, as witnessed in the engineering of  classical networks in the past decades \cite{magnus}.

One critical performance metric of  distributed algorithms is their rates of convergence.  As a foundational block for distributed algorithms, classical gossiping in randomized form achieve asymptotical convergence,  whose speed of convergence is governed by the underlying network structure\cite{Boyd2006}. Even finite-time convergence  is possible for classical gossiping under selected network topologies \cite{Shi-TON2016}. In order to improve the rate of convergence of distributed algorithms, either classical\cite{Boyd-SIAM-2006} or quantum\cite{Shi-TAC}, an immediate thought was to respect the network structure while adjusting the weights of the links representing strength of interactions. With the ability of designing the network structure, optimization is not an easy problem due to the arising  combinatorial obstacles. However, utilizing certain micro-structures   such as cliques \cite{arxiv-clique-2017}, i.e., local complete subgraphs of a network, one can resolve previously impossible convergence requirements or significantly improve convergence speed.

This paper aims to establish a framework for the acceleration of quantum gossip algorithms by introducing  clique operations to networks of qubits. For a local complete subgraph over such networks, cyclic permutations are used to define their collective interactions which can be physically realized by a series of local environments. The focus is then the convergence conditions and convergence rates with deterministic or random scheduling of the cliques. We first show that at reduced states, these cliques have the same acceleration effects, which can even enable finite-time convergence for suitable network structures. Next, we show that for random selection of cliques, the rate of convergence is improved by $\mathcal{O}(k/n)$ at reduced states, where $k$ is the size of the cliques and $n$ is the number of qubits in the network. The rate of convergence of the network coherent states  is established via   the spectrum of a mean-square error evolution matrix. Explicit calculation of such matrix seems to be extremely difficult, however,  the effect of cliques on the coherent states can be seen  via numerical examples. It is surprising to observe that  using larger quantum cliques does not necessarily accelerate  the network density aggregation. This shows that the dynamics of a quantum network is not entirely determined  by its  classical topology.

\section{Main Results}\label{sec:model}
\subsection{Open Quantum Networks}
We consider a group of quantum nodes each holding a quit indexed in the set $\mathrm{V}=\{1,2,\dots,n\}$. Time is slotted for $t=0,1,2,\dots$, and  at time $t$ the state of qubit $i$ is denoted by a density operator (matrix) $\rho_i(t)\in\mathbb{C}^{2\times 2}$. The network state is denoted by the density operator $\rho(t)\in \mathbb{C}^{2n\times 2n}$.  The qubits can be locally connected by a series of environments, which are by themselves quantum systems as well. These local environments induce a classical interaction structure which is described by a generalized  graph $\mathrm{G}=(\mathrm{V},\mathrm{E})$, where  each element in $\mathrm{E}$ is a nonempty and non-singleton subset of $\mathrm{V}$. For example, $e=\{1,2,3\}$ is a generalized edge among three nodes $1$, $2$, and $3$. We index the elements in $\mathrm{E}$ by $e_1,\dots,e_c$ for some $c>0$. The quantum generalized interaction graphs is called $k$-regular if $|e_j|=k$ for all $j=1,\dots,c$.
\begin{figure}[ht]
\centering
\includegraphics[width=4.2in]{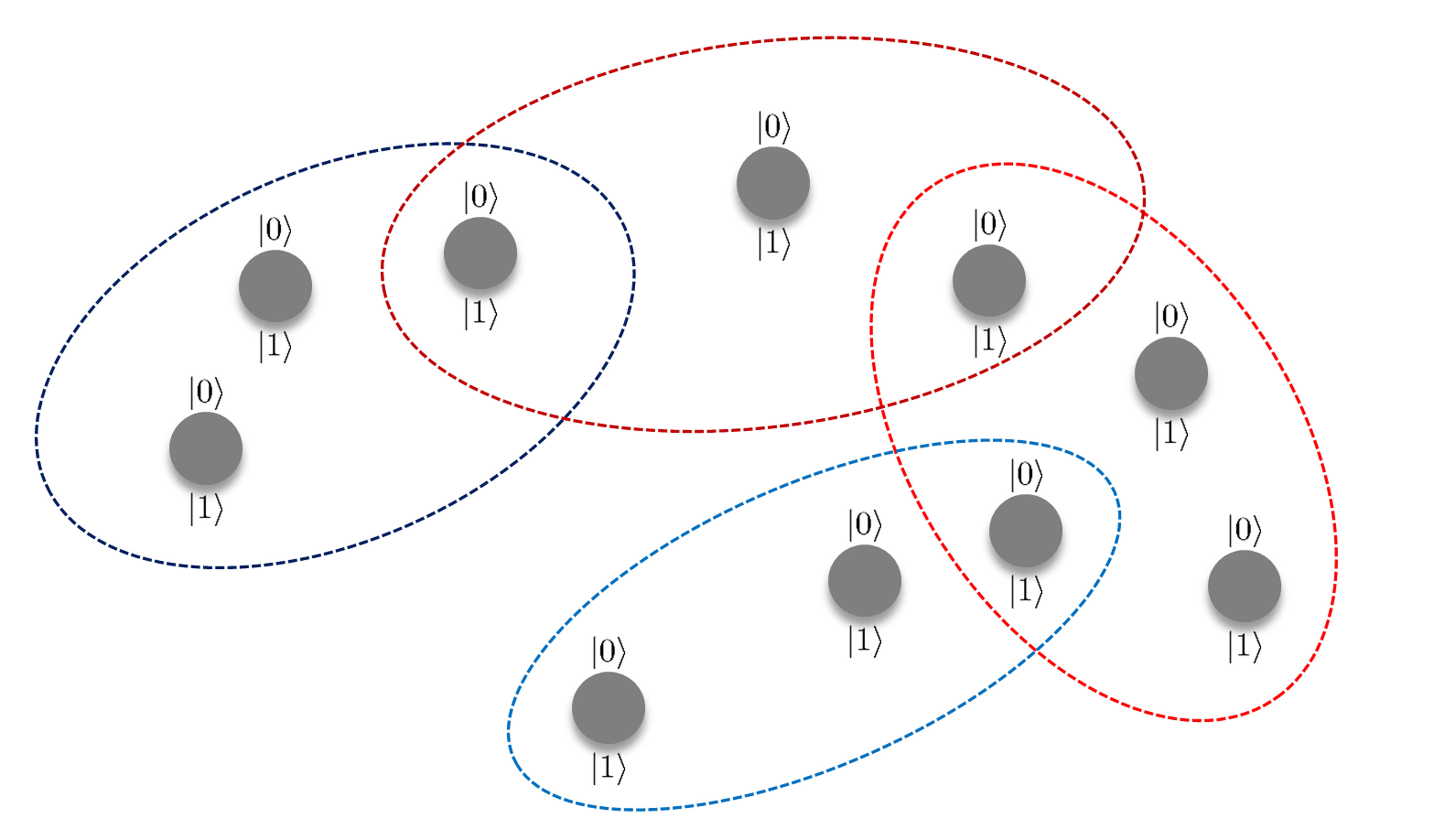}
\caption{An open quantum network interconnected by local environments. Each node holds a qubit; each dashed circle represents a local environment, where the encircled qubits forming  a quantum subsystem as a whole interact with such environment. The entirety of the system including both the qubits and the local environments forms  a closed quantum system whose state evolution is governed by unitary operations.  }
\label{fig:ori}
\end{figure}

 Recall that a permutation over a finite alphabet is a bijective mapping over the set onto itself. Particularly,  a cyclic permutation  is a permutation   which maps the elements of certain  subset  to each other in a cyclic fashion,  while mapping  each of the other elements to itself. The set of all permutations over the set $\mathrm{V}$ is called the $n$'th permutation group, denoted by $\mathbf{P}$. Associated with any $e_m$, we define a permutation  $\pi_m$ over the node set $\mathrm{V}$ in the way that \begin{itemize}
\item [(i)] $\pi_m(j)=j$ if $j\neq e_m$ for any $j=1,\dots,n$;
\item [(ii)] $\pi_m$ is a cyclic permutation.
\end{itemize}
For example with $n=4$, a permutation satisfying
$$
\pi(1)=2,\pi(2)=3,\pi(3)=1,\pi(4)=4
$$
is a cyclic permutation defined over a generalized edge $e=\{1,2,3\}$. We note that if the size of $e_m$ is $k$, the number of cyclic permutations over $e_m$ is $(k-1)!$. Here for the moment we assume that $\pi_m$  is an arbitrary  cyclic permutation to ease the presentation.  Note also that any permutation $\pi$ over $\mathrm{V}$ further induces a quantum permutation operator $U_\pi$ over the qubit network by
$$
U_\pi(|\psi_1\rangle\otimes |\psi_2\rangle\otimes \dots \otimes |\psi_n\rangle)=|\psi_{\pi(1)}\rangle\otimes |\psi_{\pi(2)}\rangle\otimes \dots \otimes |\psi_{\pi(n)}\rangle
$$
with $|\psi_i\rangle$ being any unit vector of the state space of qubit $i$.

\subsection{Deterministic Quantum Clique Gossiping}
Let $\sigma(\cdot)$ be a mapping from $\mathbb{Z}_{\geq 0}$ to  $\{1,\dots,c\}$. Practically, the mapping $\sigma(\cdot)$ selects a  multi-vertex  link from the generalized  graph $\mathrm{E}$ by assigning $e_{\sigma(t)}$ at time $t$. It is natural to assume that $\sigma(\cdot)$ is a periodic signal going through every element in the set $\mathrm{E}$. When $e_{\sigma(t)}$ is selected, the local environment associated with the qubits in $e_{\sigma(t)}$ is engineered so that  the network density operator evolves along
\begin{align}\label{sys}
\rho(t+1)=\sum_{\tau=1}^{|e_{\sigma(t)}|}U_{\pi_{\sigma(t)}^\tau}^\dag \rho(t) U_{\pi_{\sigma(t)}^\tau}/|e_{\sigma(t)}|.
\end{align}
This defines a deterministic quantum clique-gossiping algorithm.  Clearly, when each $e_{\sigma(t)}$ contains  only  two nodes, this quantum clique-gossiping becomes a standard quantum gossip algorithm. It should be pointed out that the realization of such a discrete-time quantum algorithm can be made through open quantum systems \cite{Nielsen,rivas2011open}, where the state evolution of the qubits is in continuous time but by switching the dissipative operators the algorithm (\ref{sys}) is achieved in an approximate sense along the switching instants.    The following result holds.
\begin{theorem}\label{thm1}Along the deterministic quantum clique gossip algorithm (\ref{sys}), the following statements hold.

\noindent (i) The network achieves  reduced-state consensus   in the sense that
$$
\lim_{t\to \infty}\rho_i(t)=\sum_{j=1}^n \rho_j(0)/n
$$
if and only if $\bigcup_{j=1}^c e_j=\mathrm{V}$ and  $e_m \bigcap \big(\bigcup_{j=1,j\neq m} e_j \big)\neq \emptyset$.

\noindent (ii)
The network density operator $\rho(t)$ satisfies
$$
\lim_{t\to \infty}\rho(t)=\sum_{\pi\in\mathbf{S}_{\mathrm{E}}}U^\dag_\pi \rho(0) U_\pi/|\mathbf{S}_{\mathrm{E}}|
$$
where $\mathbf{S}_{\mathrm{E}}$ is the generating subgroup by the permutations in the set $\{\pi_1,\dots,\pi_c\}$ associated with the $e_j\in \mathrm{E},j=1,\dots,c$.
\end{theorem}
This result suggests that as long as all the local environments cover the whole qubit network with sufficient connectivity, the reduced states of the qubits will asymptotically reach an average consensus. However, this condition is not enough for the symmetric-state consensus since the local environment-induced quantum evolution possesses  invariant subspaces along the coherent states, which prevents a fully symmetric mixing of the quantum states. Moreover, we would like to point out that the two convergence results   illustrated in the above theorem are both at exponential rate, consistent with the results under continuous dynamics\cite{Shi-TAC}. The following result shows the possibility of using $k$-regular interaction graphs to ensure reduced-state convergence in some finite time steps.

\begin{theorem}\label{thm2}
There exists a $k$-regular interaction graph $\mathrm{G}$  under which a quantum clique-gossiping algorithm (\ref{sys}) can   converge to a reduced-state consensus in some finite time steps   if and only if $n$ is divisible by $k$ with the same prime factors as $k$.
\end{theorem}
Particularly, if there exist factorizations $k=p_1^{r_1}\cdots p_d^{r_d}$ and $n=p_1^{s_1}\cdots p_d^{s_d}$ with $p_1,\dots,p_d$ being prime numbers and $s_i\geq r_i>0$ for all $1\leq i\leq d$, then  a fastest  $k$-regular quantum clique-gossiping algorithm  drives the qubit states to
$$
\rho_i(T)=\sum_{j=1}^n\rho_j(0)/n
$$ in
        $
       T=n \Big(\max_{1\leq i\leq d}\left\lceil \frac{s_i}{r_i}\right\rceil\Big)/k
        $
        steps. This result illustrates that cliques have exactly the same acceleration effects at the reduced states in the quantum setting as the classical case\cite{arxiv-clique-2017}. As a matter of fact, the reduced states follow a similar type of evolution where the analysis for classical networks becomes directly applicable.
\subsection{Random Quantum Clique Gossiping}
 The cyclic permutations can also be selected in a randomised fashion. Randomization in general will provide the network with the ability of self-organizing node updates.   For random scheduling signal, not only the convergence conditions, but also  the convergence speed of the quantum states are of interest. There are indeed many possible quantum interaction graphs $\mathrm{G}=(\mathrm{V},\mathrm{E})$, and on top of that there are also many  choices of the distributions of the random selection among the generalized edges in $\mathrm{E}$. Nonetheless we can benchmark our analysis to the simple case where $\mathrm{E}$ is $k$-regular containing all possible generalized links with $k$ different vertices (this means $\mathrm{E}$ has a total of $\binom nk$ entries), and at each time step each generalized link is selected independently  with equal probability. Moreover, of the $(k-1)!$ cyclic permutations over the selected $k$-node generalized set, we also assume it is randomly selected with equal probability. The resulting permutation selection process is denoted as $\bm{\pi}(t)$. Let $\mathbf{P}_k$ contain all permutations that define a $k$-cyclic permutation over a subset of $k$ nodes and induce an identity mapping over the rest $n-k$ nodes.  Then  independent with time,  $\bm{\pi}(t)$ selects a  permutation   with equal probabilities  from $\mathbf{P}_k$, and the resulting quantum network state evolution is described by
   \begin{align}\label{sys-random}
\rho(t+1)=\sum_{\tau=1}^{k}U_{\bm{\pi}^\tau(t)}^\dag \rho(t) U_{\bm{\pi}^\tau(t)}/k.
\end{align}

For the reduced states of the qubits, the following theorem holds.

\begin{theorem}\label{thm3} Let $k\leq n-1$.
Along the algorithm (\ref{sys-random}) the reduced states  $\rho_i(t)$ converge to a consensus both almost surely and in the mean-square sense. Particularly, the convergence speed is characterized by $\nu=(n-k)/(n-1)$ in the sense that
\begin{align}
 0<\max_{\rho_1(0),\dots \rho_n(0)}\limsup_{t\to\infty}\mathbb{E} \bigg\{\sum_{i=1}^n\Big\|\rho_i(t)-\sum_{j=1}^n\rho_j(0)/n\Big\|^2\bigg\}/{\nu^t}<\infty.
\end{align}
\end{theorem}
The case with $k=n$ is trivial since the reduced states of the qubits will reach a consensus in one step deterministically. It is clear from this result that  the rate of convergence is precisely improved by
$$
\frac{n-2}{n-1}-\frac{n-k}{n-1}=\frac{k-2}{n-1}
$$
when size-$k$ cliques are used instead of standard gossiping corresponding to $k=2$. Therefore, for large $n$ approximately  $\mathcal{O}(k/n)$ is added to the rate of convergence using clique gossiping for the qubits' reduced states.   On the other hand, for the network state $\rho(t)$, we establish the following understandings.
\begin{theorem}\label{thm4}
Let $n\geq 5$. Along the algorithm (\ref{sys-random}) the following statements stand.

\noindent(i) The quantum network reaches a symmetric-state consensus  $
\lim_{t\to \infty}\rho(t)=\sum_{\pi\in\mathbf{P}}U^\dag_\pi \rho(0) U_\pi/n!
$
both almost surely and in the mean-square sense  if $k$ is an even number.

\noindent (ii) The quantum network reaches  $
\lim_{t\to \infty}\rho(t)=2\sum_{\pi\in\mathbf{P}_{\rm even}}U^\dag_\pi \rho(0) U_\pi/n!
$
both almost surely and in the mean-square sense  if $k$ is an odd number, where $\mathbf{P}_{\rm even}$ is the subset of $\mathbf{P}$ containing  all even permutations over $\mathrm{V}$.
\end{theorem}
In the above result, either $k$ is even or odd the convergence speed is characterized by some $\nu_\ast>0$ in the form of
\begin{align}
 \limsup_{t\to\infty}\mathbb{E} \Big\{\big\|\rho(t)-\rho(\infty)\big\|^2\Big\}/{\nu_\ast^t}<\infty.
\end{align}
Introducing
\begin{align}\label{matrix}
M=\sum_{\pi\in \mathbf{P}_k}\Big(\sum_{\tau=1}^{k} U_{\pi^\tau}   \otimes U_{\pi^\tau}  /k\Big)\Big(\sum_{\tau=1}^{k} U_{\pi^\tau} ^\top \otimes U_{\pi^\tau}^\top   /k\Big)\cdot \frac{1}{(k-1)!}\cdot {\binom{n}{k}}^{-1},
\end{align}
the convergence rate $\nu_\ast$ can be described by
$$
\nu_\ast=\max_{\lambda\neq 1\in\Theta(M)}|\lambda|,
$$ where $\Theta(M)$ denotes the spectrum of the matrix $M$. An explicit calculation of $\nu_\ast$ is rather challenging due to the complexity of $M$.

\subsection{Examples}
\subsubsection{Reduced State Convergence}
\begin{figure}[ht]
\centering
\includegraphics[width=5in]{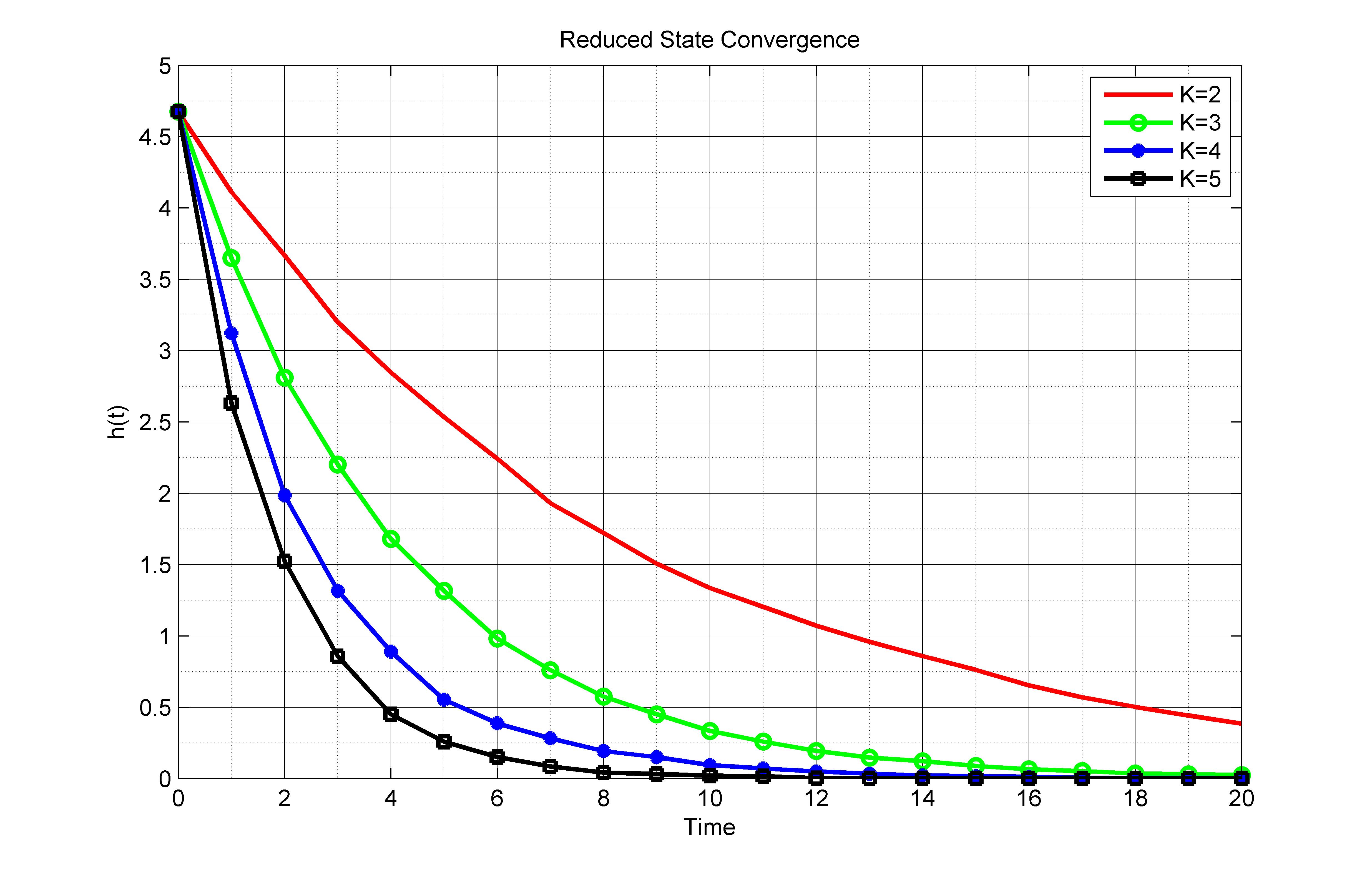}
\caption{Convergence of the qubit reduced states   for a $10$-qubit quantum network under clique gossiping with qubit cliques of size $2$, $3$, $4$, and $5$, respectively. }
\label{fig:reduced}
\end{figure}

We consider a network of $10$ qubits. Let the initial network state be $\rho(0)=A\otimes A$ with
$$
A=\big(|0\rangle \langle 0|\big)\big(|1\rangle \langle 1|\big)\big(|+\rangle \langle +|\big)\big(|-\rangle \langle -|\big)\big(|0\rangle \langle 0|\big)
$$
where $|+\rangle=\frac{1}{\sqrt{2}} \big(|0\rangle+|1\rangle \big)$ and $|-\rangle=\frac{1}{\sqrt{2}} \big(|0\rangle-|1\rangle \big)$. The clique selection follows from the random scheduling $\bm{\pi}(t)$. In Figure \ref{fig:reduced} we plot the function
$$
h(t)=\mathbb{E} \bigg\{\sum_{i=1}^n\Big\|\rho_i(t)-\sum_{j=1}^n\rho_j(0)/n\Big\|^2\bigg\}
$$
for $k=2$, $3$, $4$, and $5$, respectively. Clearly as the size of the cliques increases, the rate of convergence increases at the qubit reduced states.

\subsubsection{Network State Convergence}
Apparently it is hard to calculate the mean-square error propagation matrix $M$ through (\ref{matrix}). For a network with $5$ qubits with initial state
$
\rho(0)=A,
$
in Figure 3 we plot the function
$$
g(t)=\mathbb{E} \Big\{\big\|\rho(t)-\rho(\infty)\big\|^2\Big\}
$$
for $k=2$, $3$,  and $4$, respectively, where $\rho(\infty)$ area calculated from the results of Theorem \ref{thm4}. As one can see the rate of convergence for the network density operator no longer monotonically depends on the size of the local quantum cliques that are used. This is in contrast to the reduced-state evolution, suggesting nonlinear dependence of the quantum network state
dynamics and the classical network structure.
\begin{figure}[H]
\centering
\includegraphics[width=5in]{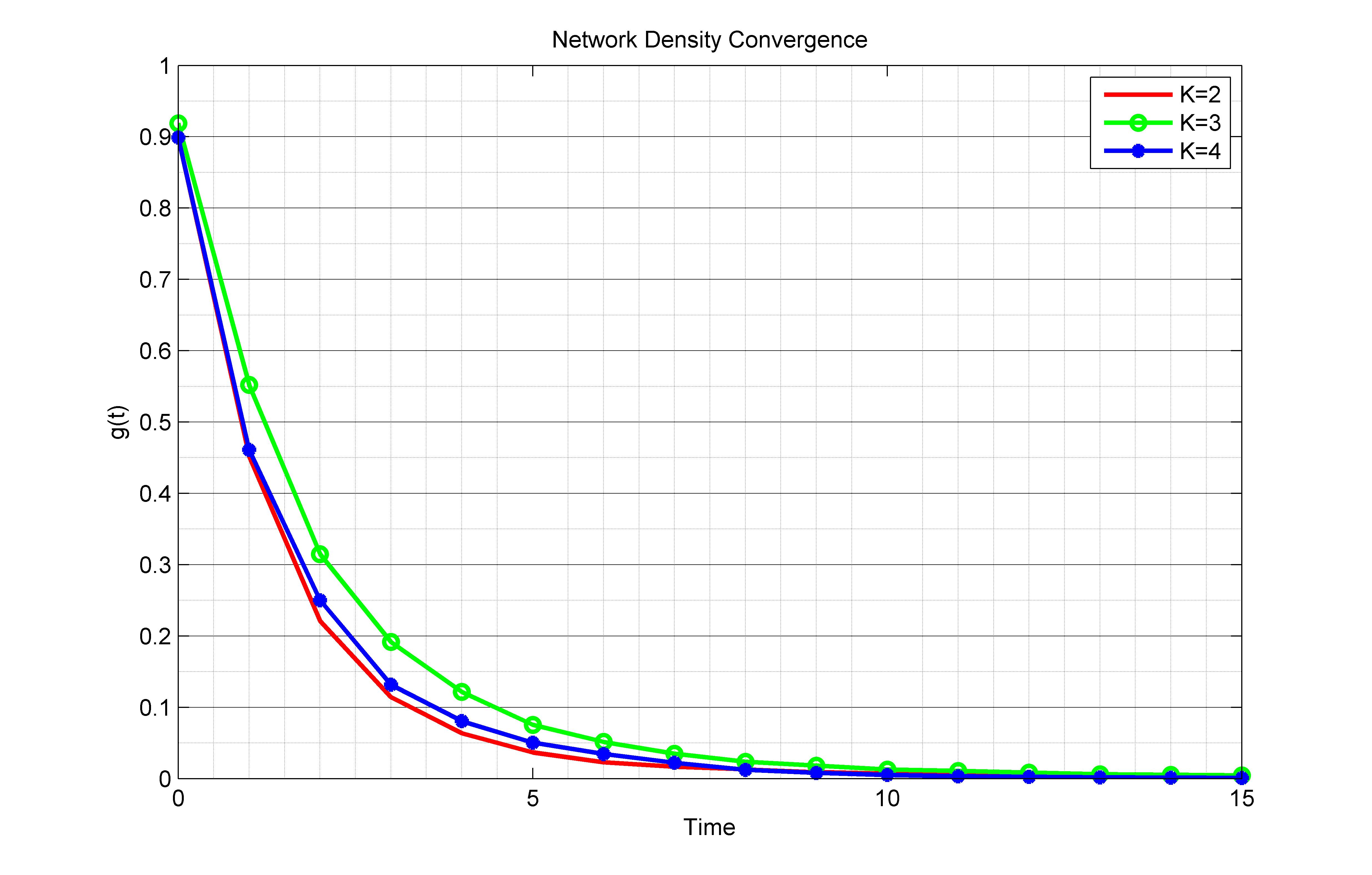}
\caption{Convergence of the network density operator  for a $5$-qubit quantum network under clique gossiping with qubit cliques of size $2$, $3$, and $4$, respectively. }
\label{fig:density}
\end{figure}

\section{Methods}
\subsection{Proof of Theorem \ref{thm1}}
(i) Let $\mathcal{H}_k$ be the two-dimensional Hilbert space associated with the $k$th qubit. Then clearly $\rho_k(t)={\rm Tr}_{\otimes_{j\neq k} \mathcal{H}_j}(\rho(t))$ as the reduced state of qubit $k$ after tracing out the rest of the qubits.  From the algorithm (\ref{sys}) it is clear that at the reduced states there holds
\begin{equation}\label{sys-reduced}
\rho_i(t+1)=\begin{cases}
 \sum_{j\in e_{\sigma(t)}}\rho_j(t)/|e_{\sigma(t)}|     & \quad \text{if}\  i\in  e_{\sigma(t)}; \\
\rho_i(t) & \quad \text{otherwise.}
  \end{cases}
\end{equation}
This  defines a matrix-valued classical averaging consensus algorithm\cite{magnus}, where the updates happen along each entry of the matrices independently.  From the results of\cite{Jadbabaie2003} we know that the desired convergence holds if and only if the edge set
$$
\Big\{ \{i,j\}: i,j\in e_k\ \text{for some}\ k=1,\dots,c\Big\}
$$
forms a connected undirected  graph over the node set $\mathrm{V}$, which is in turn equivalent to the conditions that  $\bigcup_{j=1}^c e_j=\mathrm{V}$ and  $e_m \bigcap \big(\bigcup_{j=1,j\neq m} e_j \big)\neq \emptyset$.

\noindent(ii) This part of the conclusions is  analogous to the results for continuous-time quantum consensus dynamics\cite{tac-2017}. First of all, from vectoring the network density operator $\rho(t)$ we know that the algorithm (\ref{sys}) defines a converging sequence by applying the Perron-Frobenius theory on the state transition matrix \cite{tac-2017}. Next, the convergence limit of $\rho(t)$ must be invariant under any permutation $U_{\pi_k}$ associated to any $e_k\in \mathrm{E}$. This leaves
$ \sum_{\pi\in\mathbf{S}_{\mathrm{E}}}U^\dag_\pi \rho(0) U_\pi/|\mathbf{S}_{\mathrm{E}}|
$ being the consensus limit as the only possibility.

We have now completed the proof of the theorem.

\subsection{Proof of Theorem \ref{thm2}}
The reduced-state representation (\ref{sys-reduced}) of the algorithm  is a classical clique-gossip algorithm\cite{arxiv-clique-2017}. Therefore, the desired convergence possibility and complexity results follow readily from Theorem 3 of \cite{arxiv-clique-2017}. This is consistent with the physical intuition that the reduced states at each qubit of the quantum network define classical quantities.

\subsection{Proof of Theorem \ref{thm3}}
We stack the reduced state $\rho_i(t)$ into
$$
\bm{\rho}(t)=
\begin{bmatrix}
    \rho_1(t)\\ \rho_2(t) \\ \vdots\\ \rho_n(t)
\end{bmatrix}
$$
Then there holds from the random quantum clique gossip algorithm that
\begin{align}
\bm{\rho}(t+1)= \big(\mathsf{W}_t\otimes I_2\big)\bm{\rho}(t)
\end{align}
where  $\mathsf{W}_t$ is a random matrix and $I_m$ is the $m\times m$ identity matrix.

Let $e_i$ be the unit $n$-dimensional  vector with the $i$th entry being $1$. Introduce
$$
W_{i_1\dots i_k}=\frac{1}{k}(e_{i_1}+\dots e_{i_k})(e_{i_1}+\dots e_{i_k})^\top+\sum_{j\notin\{i_1,\dots,i_k\}}e_je_j^\top
$$
for any $i_1,\dots,i_k\in\mathrm{V}$ with the $i_j$ being pairwise distinct. We can verify by direct calculation that each $W_{i_1\dots i_k}$ satisfies
$$
W_{i_1\dots i_k}=W_{i_1\dots i_k}^\top,\ \ W_{i_1\dots i_k}^2=W_{i_1\dots i_k}.
$$ Now that  $\mathrm{E}$ contains all generalized edges with size $k$ and $\sigma(t)$ selects each edge with equal probability independent with time, the distribution of $\mathsf{W}_t$ can be written as
 \begin{align}
 \mathbb{P}\Big(\mathsf{W}_t= W_{i_1\dots i_k}\Big)=1/\binom{n}{k}
 \end{align}
for any $W_{i_1\dots i_k}$. Consequently, we obtain
$$
\mathbb{E}\big\{\mathsf{W}_t^\top\mathsf{W}_t \big\}=\mathbb{E}\big\{\mathsf{W}_t\big\}=\frac{k-1}{n(n-1)}\mathbf{1}\mathbf{1}^\top+\Big(\frac{n-k+1}{n} -\frac{k-1}{n(n-1)} \Big)I_n,
$$
which implies that $\mathbb{E}\big\{\mathsf{W}_t^\top\mathsf{W}_t \big\}$ has an eigenvalue $(n-k)/(n-1)$ with multiplicity $n-1$ and another eigenvalue $1$ with multiplicity one. Invoking the analysis for standard gossip algorithm\cite{Boyd2006}, there holds for $$
\nu=\lambda_{\rm max}  \Big( \mathbb{E}\big\{\mathsf{W}_t^\top\mathsf{W}_t \big\}-\mathbf{1}\mathbf{1}^\top\Big)=(n-k)/(n-1)
$$
that
\begin{align*}
 \max_{\rho_1(0),\dots, \rho_n(0)}\limsup_{t\to\infty}\mathbb{E} \bigg\{\Big\|\rho_i(t)-\sum_{j=1}^n\rho_j(0)/n\Big\|^2\bigg\}/{\nu^t}<\infty.
\end{align*}

This implies that the $\rho_i(t)$ converges to reduced-state consensus in the mean-square sense. Since such mean-square convergence is exponential, almost sure convergence is also guaranteed. We have now completed the proof of the desired result.

\subsection{Proof of Theorem \ref{thm4}}
Recall that $\mathbf{P}_k$ is the set containing  all permutations that define a $k$-cyclic permutation over a subset of $k$ nodes and induce an identity mapping over the rest $n-k$ nodes. From (\ref{sys-random}) we can vectorize the network density operator by
$X(t)={\rm vec}(\rho(t))\in\mathbb{C}^{4^n}$ and obtain
\begin{align} \label{random}
X(t+1)=\Big(\sum_{\tau=1}^{k} U_{\bm{\pi}^\tau(t)} ^\top \otimes U_{\bm{\pi}^\tau(t)}^\dag   /k\Big) X(t)\nonumber\\
=\Big(\sum_{\tau=1}^{k} U_{\bm{\pi}^\tau(t)} ^\top \otimes U_{\bm{\pi}^\tau(t)}^\top   /k\Big) X(t),
\end{align}
where in the second equality we have used the fact that $U_{\bm{\pi}^\tau(t)}$ is a real matrix. Under this vectorization the dynamics of $X(t)$ defines a classical consensus dynamics\cite{Shi-TAC}. Convergence in both mean-square and almost sure sense becomes immediate following the same argument as used in \cite{tac-2017} and the proof of Theorem \ref{thm1}.(ii), while the consensus limit would be
$$
\rho(\infty)=\sum_{\pi\in\mathbf{S}_k} U_\pi \rho(0)U_\pi/|\mathbf{S}_k|
$$
where $\mathbf{S}_k$ is the generating subgroup from the set $\mathbf{P}_k$. It is easy to verify that $\mathbf{S}_k$ is a normal subgroup of $\mathbf{P}$. We now investigate two cases, respectively.
\begin{itemize}

\item Suppose $k$ is an odd number. Then clearly $\mathbf{S}_k$ is a subgroup of $\mathbf{P}_{\rm even}$. While it is well known that for $n\geq 5$, $\mathbf{P}_{\rm even}$ contains no proper normal subgroup. Therefore,  $\mathbf{S}_k$ must be $\mathbf{P}_{\rm even}$.

\item Suppose $k$ is an even number. Then $\mathbf{S}_k\cap \mathbf{P}_{\rm even}$ is a normal subgroup of $\mathbf{P}_{\rm even}$. Thus, $\mathbf{S}_k\bigcap \mathbf{P}_{\rm even} = \mathbf{P}_{\rm even}$. As $\mathbf{S}_k$ contains elements that are not in $\mathbf{P}_{\rm even}$, it must hold that $\mathbf{S}_k=\mathbf{P}$.
\end{itemize}
This proves the convergence statements.  For the rate of convergence, applying the analysis in the proof  of Theorem \ref{thm3} over the recursion  (\ref{random}) we know that the rate of convergence $\nu_\ast$ is given by
$$
\nu_\ast =\mathbb{E}\big\{ \mathsf{M}_t^\top \mathsf{M}_t\big\}
$$
where
$
\mathsf{M}_t=\Big(\sum_{\tau=1}^{k} U_{\bm{\pi}^\tau(t)} ^\top \otimes U_{\bm{\pi}^\tau(t)}^\top   /k\Big)$ is a random matrix. The form of $\nu_\ast$ can be made clear when we pick up the distribution of $\mathsf{M}_t$ from the distribution of $\bm{\pi}(t)$, which is exactly the equation of $M$ in (\ref{matrix}).

\section{Conclusions}
We have established a framework for the acceleration of quantum gossip algorithms by introducing  clique operations based on cyclic permutations. It was shown that at reduced states, the cliques have the same acceleration effects as   in accelerating classical gossip algorithms, under which  finite-time convergence is achievable for suitable network structures. For randomized selection of cliques, it was proven  that the rate of convergence is precisely improved by $\mathcal{O}(k/n)$ at reduced states, where $k$ is the size of the cliques and $n$ is the number of qubits in the network. It remains unanswered regarding how precisely cliques would affect the dynamics of the coherent states of the entire qubit network, where unique quantum features such as entanglements lie in. That would be a natural future direction for the study of distribute quantum algorithms.

\end{document}